\documentclass[pra]{revtex4}
\usepackage{amssymb}
\usepackage{amsmath}

\newcommand{\vc}[1]{\boldsymbol{#1}}

\begin{document}

\title{Spin Force and Torque in Non-relativistic Dirac Oscillator on a Sphere}

\author{M. S. Shikakhwa}
\affiliation{Physics Group, Middle East Technical University Northern Cyprus Campus,\\
Kalkanl\i, G\"{u}zelyurt, via Mersin 10, Turkey}

\begin{abstract}
 The spin force operator on a non-relativistic Dirac oscillator ( in the non-relativistic limit the Dirac oscillator is a spin one-half 3D harmonic oscillator with strong spin-orbit interaction)  is derived using the Heisenberg equations of motion and is seen to be formally similar to the force by the electromagnetic field on a moving charged particle . When confined to a sphere of radius R, it is shown that the Hamiltonian of this non-relativistic oscillator can be expressed  as a mere kinetic energy operator with an anomalous part. As a result, the power by the spin force and  torque operators in this case are seen to vanish. The spin force operator on the sphere is calculated explicitly and  its torque is shown to be equal to the rate of change of the kinetic orbital angular momentum operator, again with an anomalous part. This, along with the conservation of the total angular momentum, suggest that the spin force exerts a spin-dependent torque on the kinetic orbital angular momentum operator in order to conserve total angular momentum. The presence of an anomalous spin part in the kinetic orbital angular momentum operator gives rise to an oscillatory behavior  similar to the \textit{Zitterbewegung}. It is suggested that the underlying physics that gives rise to the spin force and the \textit{Zitterbewegung} is one and the same in NRDO and in systems that manifest spin Hall effect.
\end{abstract}

\maketitle

\section{Introduction}
The  spin force on spin one-half particles whose dynamics is governed by a Hamiltonian with spin-orbit interaction (SOI) has been the subject of interest over the last decade. The main motivation was the fact that this spin force was believed to provide an understanding of the spin Hall effect (SHE) an interesting phenomenon that was predicted  and later observed  to take place in some planar systems subject to SOI \cite{Dyakonov,Hirsch,Mukarami,Bernevig,Sinova,Bakun,Valenzuela},  mainly those arising from the structure inversion asymmetry known as the Rashba SOI (RSOI) \cite{rashba}, or  bulk inversion asymmetry; the Dresselhauss SOI (DSOI)\cite{dresselhaus}. Shen \cite{Shen} has demonstrated that a planar spin one-half particle subject to a linear and constant SOI experiences a spin-dependent transverse force that is proportional to the square of the electric field giving rise to the SOI. He suggested that this force could explain the separation of spins with different polarizations; an essential precondition for the SHE. He also noted that the \textit{Zitterbewegung} behavior in such systems can be attributed to this force. Tan and Jalil \cite{Tan}, working within the gauge field formalism \cite{Anandan, Frohlich, Jin, shikakhwa12} for the SOI, introduced an intuitive picture for this spin force as the force by the field curvature of this gauge field that coupled to the particle through its spin degree of freedom. In the work \cite{Ho} the spin force picture was employed to calculate the  SHE conductivity. In all these works the consideration was naturally limited to planar systems and the linear momentum of the particle was a constant of the motion.\\
The model known as the Dirac Oscillator (DO) first introduced by It\^{o} \textit{et. al.} \cite{Ito}  but became famous after its rediscovery by Moshinsky and Szczepaniak \cite{Moshinsky} was introduced in order to have a Dirac equation which is linear in both the momentum and position operators. This was achieved by introducing a non-minimal coupling to a linear potential via the substitution $\vc{p}\rightarrow \vc{p}-im\omega\vc{r}\beta$ resulting in a Dirac equation of the form:
\begin{equation}\label{DO relativistic}
i\hbar\partial_t\psi=[c\vc{\alpha}\cdot(\vc{p}-im\omega\vc{r}\beta)+mc^2\beta]\psi
\end{equation}
where $\vc{\alpha}=\beta\vc{\gamma},\beta=\gamma_0$ and the $\gamma$'s are the Dirac matrices.
In the work \cite{Brito} it was suggested that this linear term can be viewed as the non-minimal coupling of a spin one-half neutral particle to the electric field of a uniformly charged sphere, thus providing a physical interpretation of the linear interaction introduced by hand. The DO found applications in many areas including mathematical physics, nuclear and subnuclear physics and optics \cite{Quesne} (see also the review \cite{Sadurni} and the references therein).
The non-relativistic limit  of Eq.(\ref{DO relativistic}) gives the Schr\"{o}dinger equation \cite{Moshinsky}:
\begin{equation}\label{NRDO}
H\psi=\left(\frac{\vc{p}^{2}}{2m}+\frac{1}{2}m\omega^2r^2-\frac{3}{2}\hbar\omega-2\frac{\omega}{\hbar}\vc{L}\cdot\vc{S}\right)\psi=E\psi
\end{equation}
where $\vc{p}=-i\hbar\nabla$;$\vc{L}=\vc{r}\wedge\vc{p}$ is the orbital angular momentum; $\vc{S}=\frac{\hbar}{2}\vc{\sigma}$ and $\sigma_i; i=1..3$ are the Pauli matrices. The above (NRDO) is just the Schrodinger equation of a spin one-half neutral particle subject to a 3D harmonic oscillator potential and a strong SOI term resulting from a linearly growing radial electric field. Therefore, we expect to have the spin force in action here, too, just as in the case of planar systems. The present work is an attempt to calculate and interpret the spin force acting on this particle following - generally- the same lines used in calculating the same for planar systems. We will focus more on the case when the NRDO is placed on a sphere of radius $R$ as the spin force physics in this case is more transparent on the one hand, and such a geometry is closer to condensed matter systems on the other hand. Evidently, with the DO being  just a theoretical model so far, the investigation of the spin force for this system is a theoretical curiosity. However, being a bound state problem, in contrast to the condensed matter 2D systems with SOI for which spin force is investigated, gives it an added intrinsic importance. On the other hand, as will become clear in the article, the physical interpretation of the spin force is so clear and transparent in this problem which might contribute to a better understanding of the 2D systems that have promising practical applications.

\section{Force Operator in the Non-Relativistic Dirac Oscillator}
In this section we derive the spin force operator for the 3D NRDO starting from Eq.(\ref{NRDO}) using the Heisenberg equations of motion following Anandan \cite{Anandan}. The SOI term in that equation  can be expressed as:
\begin{equation}\label{reducing so}
-2\frac{\omega}{\hbar}\vc{L}\cdot\vc{S}=-\omega\vc{p}\cdot(\vc{\sigma}\wedge\vc{r})=-\vc{p}\cdot(\vc{\sigma}\wedge\vc{E})
\end{equation}
where $\vc{E}=\omega \vc{r}$ and we have noted that $\nabla\wedge\vc{r}=0$. We can then, upon defining the $SU(2)$ gauge field $g\vc{W}=g\vec{W}^a\sigma^a; a=1..3$ as
\begin{equation}\label{W}
g\vc{W}=m\omega(\vc{\sigma}\wedge\vc{r})=m(\vc{\sigma}\wedge\vc{E});\quad g=m\omega
\end{equation}
express the  NRDO Hamiltonian as that of a neutral particle coupled to a gauge field;
\begin{eqnarray}\label{H in termsof W}
   H(\vc{W}) &=&\frac{1}{2m}(\vc{p}-g\vc{W})^2-\frac{3}{2}\hbar\omega-\frac{g^2}{2m}\vc{W}\cdot\vc{W} \\\nonumber
   &=&  \frac{1}{2m}(\vc{p}-m\omega(\vc{\sigma}\wedge\vc{r}))^2-\frac{3}{2}\hbar\omega-\frac{1}{2}m\omega^2r^2
\end{eqnarray}
The Heisenberg equations give for  the velocity operator :
\begin{equation}\label{velocity op}
\vc{v}=\frac{d\vc{r}}{dt}=\frac{[\vc{r},H]}{i\hbar}=\frac{1}{m}(\vc{p}-g\vc{W})=\frac{1}{m}(\vc{p}-m\omega(\vc{\sigma}\wedge\vc{r}))=\frac{\vc{\pi}}{m}
\end{equation}
where we have introduced the kinematic momentum $\vc{\pi}$. The term $- \frac{g}{m}\vc{W}= -\omega(\vc{\sigma}\wedge\vc{r})$ is what is known in the literature as the anomalous velocity \cite{Karplus,Chang}. Thus, we express the Hamiltonian as
\begin{eqnarray}
  H &=& \frac{m}{2}v^2 -\frac{3}{2}\hbar\omega-\frac{1}{2}m\omega^2r^2\\\nonumber
   &=& \frac{\pi^2}{2m}-\frac{3}{2}\hbar\omega-\frac{1}{2}m\omega^2r^2
\end{eqnarray}
The calculation of the force operator then reduces mainly to the calculation of the commutator $[v_i,v_j]$;
\begin{equation}\label{force operator}
F_i=\frac{d\pi_i}{dt}=\frac{m^2}{2i\hbar}([v_i,v_j]v_j+v_j[v_i,v_j])
\end{equation}
where we have ignored the contribution of the potential to the force as this merely provides the Hooke's law force which is spin-independent.The commutator turns out to be
proportional to the field strength tensor $\frac{m^2}{2i\hbar}[v_i,v_j]=G_{ij}^c\sigma^c$.  So, the force operators reads formally
\begin{equation}\label{force formaly}
F_i=(G_{ij}^c\sigma^cv_j+v_j\sigma^cG_{ij})
\end{equation}
with $G_{ij}^c$ given explicitly  as:
\begin{eqnarray}\label{G}
  G_{ij}^c &=& m\omega\left(\delta^{ac}(\partial_iW_j^a-\partial_jW_i^a)+\frac{2g}{\hbar}\epsilon_{abc}W_i^aW_j^b)\right) \\\nonumber
   &=& \epsilon_{ija}(m\omega\delta^{ac}+\frac{m^2}{\hbar}E_aE_c)
\end{eqnarray}
The above expression for the force is formally  similar to the standard magnetic force on a charged particle moving with velocity $\vc{v}$. This result is a general one for any linear SOI \cite{Anandan}.
Using the above explicit expression for the field strength tensor the force is given explicitly as ($\{.,.\}$ denotes the anti-commutator) :
\begin{equation}\label{F explicit}
\vc{F}=m\omega(\vc{v}\wedge\vc{\sigma}-\vc{\sigma}\wedge\vc{v})-\frac{m^2}{\hbar}\{\vc{E}\cdot\vc{\sigma},\vc{E}\wedge\vc{v}\}
\end{equation}
The second term in the above expression of the force was obtained  by Shen \cite{Shen} for a general two dimensional hetrostructure with constant linear SOI. It was argued that it is the spin force that leads to spin separation and so the SHE. Here, $\vc{E}$ is radial and not constant. A straightforward but delicate calculation allows us to bring the force into the form:
\begin{equation}\label{F in terms of j's}
\vc{F}=-m(1+\frac{mE}{\hbar r^2})E\vc{o}^r+\frac{4E}{\hbar r}\vc{\hat{r}}\vc{S}\cdot\vc{K}+\frac{2i}{r^2}\vc{E}\wedge\vc{S}+\frac{m}{r}\vc{E}\wedge\vec{J}_r^H
\end{equation}
In the above expression $\vc{K}$ is the kinematic orbital angular momentum
\begin{equation}\label{K}
\vc{K}\equiv\vc{r}\wedge\vc{\pi}=\vc{L}-m\omega r^2\vc{\sigma}'
\end{equation}
 where $\vc{\sigma}'\equiv \vc{\hat{\theta}}\sigma_{\theta}+\vc{\hat{\phi}}\sigma_{\phi}$ and $\sigma_{\theta}= \vc{\sigma}\cdot\vc{\hat{\theta}}$...etc.; $\vc{o}^r$ is the $\vc{\hat{r}}-$polarized orbital spin current:
 \begin{equation}\label{orbital current}
\vc{o}^r\equiv \{\sigma_r,\frac{\vc{L}}{mr^2}\}
 \end{equation}
and  $\vec{J}_r^H$ - we use the arrow rather than bold font to indicate that it is a vector in spin space not the coordinate space- is the radially-convicted Hermitian spin current;
\begin{equation}\label{J Hermitian}
\vec{J}_r^H\equiv \{\vc{\sigma},p_r^H\}
\end{equation}
 with $p_r^H\equiv-i\hbar(\partial_r+\frac{1}{r})$ being the Hermitian radial  momentum \cite{shikakhwa pla ejp}, sometimes called the geometrical momentum \cite{Liu1, Liu2, Liu3, Liu4}. Expressing $\vec{J}_r^H$ in terms of this Hermitian operator renders it Hermitian. The part quadratic in the electric field in the first term of  Eq.(\ref{F in terms of j's}) is the orbital counterpart of the expression for the force obtained by Shen \cite{Shen} and declared as the main result of his work. The part linear in the electric field was absent from his expression since he used a constant SOI in contrast to our case where we have a linearly growing electric field.  One might expect the appearance of $\vc{O}^r\equiv \{\sigma_r,\frac{\vc{K}}{mr^2}\}$ instead of $\vc{o}^r$ in the first term of Eq.(\ref{F in terms of j's}). It turns out that that $\vc{O}^r=\vc{o}^r$. Finally, we note that the first and last terms in the same equation are Hermitians as they stand. As for the second and third terms, they are Hermitian together not separately. This is because the curvilinear unit vectors are not constants. We now move to consider the NRDO on a sphere.

\section{Spin Force on the Oscillator on a Sphere }

 To construct the Hermitian Hamiltonian on a sphere of radius $R$, we start with the 3D Hamiltonian, Eq.(\ref{H in termsof W}), and , as was shown in \cite{shikakhwa pla ejp}, set $p_r^H$ to zero  and $r\rightarrow R$. While for a sphere, and for this particular Hamiltonian, the result is the same as the one obtained by simply setting $\partial_r$ to zero and $r\rightarrow R$, this last approach is not correct in general \cite{shikakhwa pla ejp}. The Hamiltonian on the sphere thus reads:

 \begin{eqnarray}\label{H sphere}
  H_{sph} &=& \frac{-\hbar^2}{2m}\nabla'^2+i\hbar\omega(\sigma_\phi\partial_\theta-\sigma_\theta\frac{\partial_\phi}{\sin\theta})+\frac{1}{2}m\omega^2R^2-\frac{3}{2}\hbar\omega \\\nonumber
    &=& \frac{L^2}{2mR^2}-2(\frac{\omega}{\hbar})\vc{L}\cdot\vc{S}+\frac{1}{2}m\omega^2R^2-\frac{3}{2}\hbar\omega
 \end{eqnarray}
 where we have defined the primed  operators $\nabla'$ and $\nabla'^2$ to refer, respectively, to the gradient and Laplacian operators on the surface of the sphere;
 \begin{equation}\label{gradient and laplacian on sphere}
 \nabla'\equiv \frac{1}{R}\vc{\hat{\theta}}\partial_\theta+\vc{\hat{\phi}}\frac{\partial_\phi}{R\sin\theta},\quad \nabla'^2\equiv \frac{1}{R^2\sin\theta}\partial_\theta(\sin\theta\partial_\theta)+\frac{1}{R^2\sin^2\theta}\partial^2_\phi
 \end{equation}
  The above Hamiltonian is that  of a rigid rotor plus a strong SOI. The velocity operator can be found by calculating the time derivative of the position operator on the sphere  $\vc{R}=R\vc{\hat{r}}$ using the Heisenberg equations of motion and is just, as expected, the kinematic momentum $\vc{\pi'}$ on the sphere divided by the mass:
  \begin{equation}\label{v on a sphere}
  \vc{v'_H}=\frac{d\vc{R}}{dt}=\frac{[\vc{R},H_{sph}]}{i\hbar}=\frac{1}{m}(\vc{p'_H}-g\vc{W})=\frac{\vc{\pi'}}{m}
  \end{equation}
  where we have introduced the Hermitian canonical momentum on the sphere
 \begin{equation}\label{Hermitian v0}
\vc{p'_H}=-i\hbar(\nabla'-\frac{\vc{\hat{r}}}{R})
\end{equation}
The radial part of the above operator is crucial to guarantee Hermiticity. This fact seems a general result for the Hermitian momentum on any 2D surface embedded in 3D by confining along the coordinate normal to the surface. The details will be published elsewhere. We mention here that a similar conclusion  was obtained in the work \cite{Liu5,Liu6} by a completely different approach. The explicit form for the Hermitian velocity operator reads:
\begin{equation}\label{explicit v hermitian}
\vc{v'_H}=\frac{\vc{\pi'}}{m}=\frac{-i\hbar}{Rm}(\vc{\hat{\theta}}(\partial_\theta-\frac{im\omega R^2}{\hbar}\sigma_\phi)+\vc{\hat{\phi}}(\frac{\partial_\phi}{R\sin\theta}+\frac{im\omega R^2}{\hbar}\sigma_\theta)-\vc{\hat{r}})
\end{equation}
Therefore, the Hamiltonian on a sphere, Eq.(\ref{H sphere}),  can be written as:
\begin{equation}\label{H sphere in terms of v}
  H_{sph} = m\frac{v'^2_H}{2}-\frac{1}{2}m\omega^2R^2-\frac{3}{2}\hbar\omega-\frac{\hbar^2}{mR^2} \\
\end{equation}
We can go further and express the above Hamiltonian in terms of  the kinematic orbital angular momentum operator $\vc{K}$ defined earlier; Eq.(\ref{K}), which on the sphere assumes the form:
\begin{equation}\label{K on a sphere}
\vc{K}\equiv\vc{R}\wedge\vc{\pi'}=\vc{R}\wedge(\vc{p'_H}-g\vc{W})=\vc{L}-m\omega R^2\vc{\sigma'}
\end{equation}
with $\vc{\sigma'}\equiv\vc{\sigma}-\vc{\hat{r}}\sigma_r=\vc{\hat{\theta}}\sigma_\theta+\vc{\hat{\phi}}\sigma_\phi$. The Hamiltonian on the sphere expressed in terms of $\vc{K}$ is then:
\begin{eqnarray}\label{H sphere in terms of K}
 H_{sph}&=& \frac{K^2}{2mR^2}-\frac{1}{2}m\omega^2R^2-\frac{3}{2}\hbar\omega \\\nonumber
   &=& \frac{1}{2}I\Omega^2-\frac{1}{2}m\omega^2R^2-\frac{3}{2}\hbar\omega
\end{eqnarray}
where we have introduced the kinematic angular velocity $\vc{\Omega}\equiv\frac{\vc{K}}{mR^2}$ and the moment of inertia $I\equiv mR^2$. Note that $\vc{\Omega}$ contains the anomalous angular velocity $-\omega \vc{\sigma'}$. So, the Hamiltonian on the sphere is just  kinetic energy - with an anomalous  term, however- plus constants. Looking at Eq.(\ref{H sphere in terms of K}) and the second line of Eq.(\ref{H sphere}) together, we see that the kinetic energy is the sum of the orbital kinetic energy and the SOI. In other words, the anomalous part of the kinetic energy operator $\frac{K^2}{2mR^2}$ is the SOI; $-2(\frac{\omega}{\hbar})\vc{L}\cdot\vc{S}$. Each part of the kinetic energy is conserved separately. The fact that the kinetic energy is conserved ( and so are the magnitudes of the kinetic angular momentum $K$, and  $\Omega$ ) evidently implies that there is no energy transfer to or from the system, or in other words the forces and torques acting on the system do no work. The vanishing of this work manifests as the vanishing of  the translational ($P_{trn}$) and rotational ($P_{rot}$) powers . The vanishing of translational power  follows from the conservation of $\pi'^2$;
\begin{equation}\label{no power by F}
\frac{d}{dt}\left(\frac{\vc{\pi'}\cdot\vc{\pi'}}{2m}\right)=0=\frac{1}{2m}\left(\frac{d\vc{\pi'}}{dt}\cdot\vc{\pi'}+\vc{\pi'}\cdot\frac{d\vc{\pi'}}{dt}\right)=\frac{1}{2}\left(\vc{F'}\cdot\vc{v'_H}+\vc{v'_H}\cdot\vc{F'}\right)=P_{trn}
\end{equation}
where we have identified $\vc{F'}=\frac{d\vc{\pi'}}{dt}$. As for the vanishing of the rotational power delivered by the torque, we first note the important result which can be easily obtained upon noting that $\vc{R}\wedge\vc{\pi'}=-\vc{\pi'}\wedge\vc{R}$:
\begin{equation}\label{derivative of K}
 \frac{d\vc{K}}{dt}=\frac{d}{dt}(\vc{R}\wedge\vc{\pi'})=\frac{1}{2}\left(\vc{R}\wedge\vc{F'}-\vc{F'}\wedge\vc{R}\right)\equiv\vc{\tau}
\end{equation}
where $\vc{\tau}$ is the torque.
Then it is easy to get:
\begin{equation}\label{no power by K}
\frac{d}{dt}\left(\frac{\vc{K}\cdot\vc{K}}{2mR^2}\right)=0=\frac{1}{2}\left(\frac{d\vc{K}}{dt}\cdot\frac{\vc{K}}{mR^2}+\frac{\vc{K}}{mR^2}\cdot\frac{d\vc{K}}{dt}\right)=
\frac{1}{2}\left(\vc{\tau}\cdot\vc{\Omega}+\vc{\Omega}\cdot\vc{\tau}\right)=P_{rot}
\end{equation}
with $\vc{\Omega}$ being the kinematic angular velocity defined earlier, Eq.(\ref{H sphere in terms of K}). So, we have a situation, similar to that of a particle in a magnetic field where the force exerts a torque but does not transfer energy. In our case, the velocity is the kinematic angular velocity which, however, contain  an anomalous term; the $\vc{\sigma'}$. We will move now to the explicit evaluation of the force using the Heisenberg equations of motion and taking as the Hamiltonian that on the sphere, Eq.(\ref{H sphere}). We first split the Hamiltonian and the velocity operators as follows:
\begin{equation}\label{splitting v's}
\vc{v'_H}=\vc{v'_{0H}}+\vc{v'_f}, \quad \vc{v'_{0H}}\equiv\frac{\vc{p'_H}}{m}=\frac{-i\hbar}{m}(\nabla'-\frac{\vc{\hat{r}}}{R});\quad \vc{v'_f}\equiv -\omega \vc{\sigma}\wedge \vc{R}
\end{equation}

\begin{equation}\label{splitting H sphere}
   H_{sph}=H'_0+H'_f;\quad H'_0\equiv \frac{L^2}{2mR^2};\quad H'_f\equiv -2\frac{\omega}{\hbar}\vc{L}\cdot\vc{S}+...
\end{equation}
where dots in the above equation refer to the constant terms in the Hamiltonian which we will not consider now on. The force then can be split as
\begin{equation}\label{F split}
\vc{F'}=m\frac{d\vc{v'_H}}{dt}=\frac{m}{i\hbar}[\vc{v'_H},H_{sph}]=\vc{F'_0}+\vc{F'_{spin}}
\end{equation}
where
\begin{equation}\label{F0 definition}
\vc{F'_0}\equiv \frac{m}{i\hbar}[\vc{v'_{0H}},H'_0]
\end{equation}

is the force in the absence of the SOI which is spin-independent evidently. While irrelevant to our present discussion, we give its explicit expression and will discuss its various terms elsewhere:
\begin{equation}\label{F0 explicitly}
\vc{F'_0}=-\vc{\hat{r}}\left(\frac{mv'^2_{0H}}{R}+\frac{\hbar^2}{mR^3}\right)-i\frac{\hbar}{R^2}\vc{v'_{0H}}
\end{equation}
As for $\vc{F'_{spin}}$, upon evaluating the remaining commutators in Eq.(\ref{F split}) (other than the one in Eq.(\ref{F0 definition}) )we get the expression:
\begin{equation}\label{F spin on a sphere}
\vc{F'_{spin}}=-m(1+\frac{mE}{\hbar R^2})E\vc{o}^R+\frac{4E}{\hbar R}\vc{\hat{r}}\vc{S}\cdot\vc{K}+\frac{2i}{R^2}\vc{E}\wedge\vc{S}
\end{equation}
This expression for the force is just what one obtains  from the 3D force, Eq.(\ref{F in terms of j's}), upon applying the sphere condition: Radial Hermitian momentum $p_r^H\equiv-i\hbar(\partial_r+\frac{1}{r})$ set to zero and $R\rightarrow r$, which results in dropping of the last term in that equation. Note that the dropped term is Hermitian by itself which guarantees the Hermiticity of the remaining expression as can be checked explicitly.\\
To understand the nature of the spin force, we need to look closer at the dynamics of the system governed by the Hamilonian, Eq.(\ref{H sphere in terms of K}). As is known, while $\vc{L}$ and $\vc{S}$ are not conserved separately, $\vc{J}=\vc{L}+\vc{S}$ is conserved. This means that $\frac{d\vc{L}}{dt}=-\frac{d\vc{S}}{dt}$. For our system we have:
\begin{equation}\label{dl/dt}
\frac{d\vc{L}}{dt}=-\frac{d\vc{S}}{dt}=-\frac{2\omega}{\hbar}\vc{S}\wedge\vc{L}
\end{equation}
The meaning of this, upon noting that $L^2$, $S^2$ and $\vc{L}\cdot \vc{S}$ are also constants of the motion is that the precession of the spin in the magnetic field which is a real one induced in the rest frame of the particle, $\vc{B}\sim \vc{E}\wedge\vc{v}$, is accompanied by a precession of $\vc{L}$ in a way that preserves $\vc{L}\cdot\vc{S}$. At this point it is natural to ask if we can construct an  operator that will have its time derivative equal to $-\frac{d\vc{K}}{dt}$. Indeed, it is obvious that the following operator
\begin{equation}\label{kinematic S}
\vc{\mathcal{S}}\equiv\vc{S}+2\frac{m\omega R^2}{\hbar}\vc{S'}
\end{equation}
satisfies
\begin{equation}\label{dk/dt}
-\frac{d\vc{\mathcal{S}}}{dt}=\frac{d\vc{K}}{dt}= \vc{\tau}
\end{equation}
Moreover, it is easily seen that $\vc{\mathcal{S}}+\vc{K}=\vc{L}+\vc{S}=\vc{J}$ which upon verifying that $\vc{\mathcal{S}}\cdot\vc{K}=\vc{K}\cdot\vc{\mathcal{S}}$ means that:
\begin{equation}\label{S.K}
\vc{\mathcal{S}}\cdot\vc{K}=\frac{1}{2}(J^2-\mathcal{S}^2-{K}^2)
\end{equation}
As all the operators on the r.h.s of the above equation are constants of the motion, so is $\vc{\mathcal{S}}\cdot\vc{K}$.
Since $K^2$  but not $\vc{K}$ is a constant of the motion, then with Eqs.(\ref{derivative of K})and (\ref{dk/dt}) in mind, we can say that it is the spin force that exerts torque on $\vc{K}$ to have it precessing in a manner that keeps $\vc{K}\cdot\vc{\mathcal{S}}$ constant. This provides an interpretation of this spin force. Below, we show by explicit calculation the validity of this conjecture, thus presenting the major result of this work.\\
We first calculate the torque by $\vc{F'_{spin}}$, Eq.(\ref{F spin on a sphere}). A careful calculation demonstrates that the torque of the last two terms (combined) of $\vc{F'_{spin}}$ vanishes;
\begin{equation}\label{vanishing torque }
\vc{R}\wedge(\frac{4E}{\hbar R}\vc{\hat{r}}\vc{S}\cdot\vc{K}+\frac{2i}{R^2}\vc{E}\wedge\vc{S})-(\frac{4E}{\hbar R}\vc{\hat{r}}\vc{S}\cdot\vc{K}+\frac{2i}{R^2}\vc{E}\wedge\vc{S})\wedge\vc{R}=0
\end{equation}
The torque of the remaining term is
\begin{equation}\label{torque explicitly}
\vc{\tau}=\vc{R}\wedge(-m(1+\frac{mE}{\hbar R^2})E\vc{o}^R)-(-m(1+\frac{mE}{\hbar R^2})E\vc{o}^R)\wedge\vc{R})=(1+\frac{mRE}{\hbar})mE \vc{J'^r_0}
\end{equation}
where
\begin{equation}\label{bare radial spin current }
\vc{J'^r_0}\equiv\{\sigma_r,\frac{\vc{p'_0}}{m}\}
\end{equation}
 Recalling that $\vc{p'_0}=-i\hbar\nabla'$ , the above current is  the "bare" radially-polarized spin current. The torque is proportional to and directed along the tangential canonical velocity $\vc{v_0}$ - without the anomalous term !.It worths mentioning here that the torque by $\vc{F'_0}$ vanishes,too. One more delicate calculation gives the result:
 \begin{equation}\label{derivative of K explicitly}
\frac{d\vc{K}}{dt}=(1+\frac{mRE}{\hbar})mE \vc{J'^r_0}
 \end{equation}
 which is precisely the torque found in Eq.(\ref{torque explicitly}). This, does not only prove our conjecture, but also demonstrates that it is only the first term in the spin force in Eq.(\ref{F spin on a sphere})  that exerts the  torque needed to conserve the projection of $\vc{\mathcal{S}}$ along the kinematical orbital angular momentum. This particular force is the one that has been mainly associated with the spin Hall force and the \textit{zitterbewegung} in condensed matter systems \cite{Shen}. In fact, one can see that an oscillatory behavior of $\vc{K}$ similar to  \textit{zitterbewegung} in linear motion develops here. too. Since $\vc{\sigma'}$ oscillated in time, then noting its definition, Eq.(\ref{K on a sphere}), $\vc{K}$ oscillates in time,too. Since, as was discussed above, the oscillation of $\vc{K}$ is essential to conserve total angular momentum, then we can say that this \textit{zitterbewegung}-like behavior is a must to conserve total angular momentum of the particle. This relation between \textit{zitterbewegung} and the conservation of angular momentum for some planar systems was noted in \cite{David}.

\section{Conclusions}
The spin force for the NRDO, which is a spin one-half particle subject to a harmonic oscillator potential and a strong SOI is investigated. When confined to a sphere, the Hamiltonian of the oscillator turns out to be merely  the kinetic energy operator with an anomalous part containing the spin operator. The spin force and its corresponding torque operators are derived using the Heisenberg equations of motion. it is checked that these do not transfer energy to the particle. While the magnitude ( the square) of the kinematic orbital angular momentum and thus the kinetic energy are conserved, the kinematic angular momentum vector $\vc{K}$ is not. It precesses so that its rate of change is given by the torque of spin force. This picture is similar to that of a particle in a magnetic field. The precession of the kinematic orbital angular momentum $\vc{K}$ is such that the projection of the corresponding spin operator along it is conserved; a consequence of the conservation of the total angular momentum of the particle. During this mechanism, the anomalous part of $\vc{K}$ that contains the spin operator oscillates as a result of the spin oscillation, a mechanism similar to the \textit{zitterbewegung} in linear motion. The only part of the spin force that contributes to the torque on $\vc{K}$ is found to be the counterpart of the spin transverse force that was reported in the literature of 2D systems \cite{Shen} as the force that leads to spin separation in the SHE and the \textit{zitterbewegung}. This force is non zero only when the particle's spin has a radial component, i.e. a component along the electric field giving rise to the SOI.\\
 Finally, we note that the current analysis can be extended to other systems. One example is the DO in 2D, where one can consider confining it to a ring geometry. One can also analyze other bound atate problems with SOI, like the Hydrogen atom.  Moreover, noring that the Hamiltonian of the NRDO conisdered in this work, when the SOI is expressed as a gauge field, is gauge covariant up to a constant. So, an investigation of the meaning of the gauge transformations in this case might be of interest. Also, the relation of this system to the Aharonov-Casher effect \cite{Aharonov-Casher}might worth investigation. Work in these directions is under progress.


\end{document}